\documentclass[12pt,preprint]{aastex}
 
\slugcomment{Appearing in May 2006 PASP}
 
\shorttitle{6370 {\AA} in $\eta$ Car}
\shortauthors{Martin et al.}
 
\begin{document}


\title{Variable Unidentified Emission Near 6307 {\AA} in $\eta$ Carinae\altaffilmark{1,2}}
\author{J. C. Martin\altaffilmark{3}}
\author{K. Davidson\altaffilmark{3}}
\author{F. Hamann\altaffilmark{4}}
\author{O. Stahl\altaffilmark{5}}
\author{K. Weis\altaffilmark{6,7}}
\altaffiltext{1}{This research is part of the Hubble Space Telescope
  Treasury Project for Eta Carinae, supported by grants GO-9420 and
  GO-9973 from the Space Telescope Science Institute (STScI), which is
  operated by the Association of Universities for Research in
  Astronomy, Inc., under NASA contract NAS 5-26555. }
\altaffiltext{2}{Partially based on observations obtained with UVES at
  the ESO Very Large Telescope, Paranal, Chile (proposals
  70.D-0607(a), 71.D-0168(A), and 072.D-0524(A))}
\altaffiltext{3}{School of Physics and Astronomy, University of
  Minnesota, 116 Church Street SE, Minneapolis, MN 55455;
  martin@etacar.umn.edu}
\altaffiltext{4}{Department of Astronomy, University of Florida, P.O. Box 112055, Gainesville, FL 32611}
\altaffiltext{5}{Landessternwarte Heidelberg, K\"{o}nigsstuhl, 69117 Heidelberg, Germany.}
\altaffiltext{6}{Astronomisches Institut, Ruhr-Universit\"{a}t Bochum, Universit\"{a}tsstrasse 150, D-44780 Bochum, Germany}
\altaffiltext{7}{Lise Meitner Fellow}

\begin{abstract}
We have discovered a conspicuous unidentified variable feature near
6307 {\AA} in the spectrum of $\eta$ Carinae which is spatially
unresolved from the central star and its wind ($r \la$ 200--300 AU).
It is significant for two reasons:  such prominent unidentified
lines are now rare in this object, and this feature varies
strongly and systematically.  It exhibits a combination of
characteristics which, so far as we know are unique in $\eta$
Carinae's spectrum.  It may provide insights into the recurrent
spectroscopic events and the star's long-term brightening.

\end{abstract}


\section{Introduction}

The spectrum of $\eta$ Carinae has been extensively studied and
characterized, but unexpected changes can happen during a
spectroscopic event like the one that occurred in mid-2003.  For
instance, surprisingly strong high-excitation \ion{He}{2} $\lambda$4687
briefly appeared at that time as reported by \citet{steiner},
\citet{uves} and \citet{heii}; see the last of those papers for an
analysis of its significance.  Motivated by this example, we have
examined Hubble Space Telescope (HST) data for other transient
features.  Our search yielded an undiagnosed emission feature at 6307
{\AA} with good signal-to-noise and unusual behavior.

These data were obtained with the Space Telescope Imaging Spectrograph
(STIS), whose high spatial resolution allowed us to examine the
central star itself (or rather its inner wind) separate from the bright
nearby ejecta that contaminate all ground-based spectroscopy of $\eta$
Car (Figure \ref{ejectamap}).

The 6307 {\AA} feature is spatially unresolved from the continuum
emission of the central star and its wind in the STIS CCD data.  It
does not match any of the known atomic or molecular transitions of any
species that we expect to find in the spectrum (see \citet{thackeray},
\citet{zethsonphd} (Table \ref{zethson}), and \citet{PASPv113n788}
for extensive lists of species identified in $\eta$ Car).  This is a
significant discovery because, unlike the few other noticeable lines
that have not yet been identified, the 6307 {\AA} emission varies
conspicuously.  It is obviously correlated with $\eta$ Car's 5.5-year
spectroscopic period; indeed it temporarily disappeared during the
1998 and 2003 spectroscopic events. Moreover, following the 2003 event
this feature became much stronger than it had been in the previous
cycle, possibly indicating a link with the rapid brightening and other
mysterious developments that have been superimposed on the 5.5-year
cycle since the mid-1990's \citep{davidson99a,etaphot,halpha}.
Altogether, it is unique in the small set of lines which remain
unidentified in $\eta$ Car's spectrum. 

\section{The Data}
\subsection{Spectra}
The HST/STIS spectra in this paper were obtained as part of the $\eta$
Carinae HST Treasury Project \citep{etatp} and were reduced using
a modified version of the Goddard CALSTIS reduction pipeline (Table
\ref{stisdata}).  The modified pipeline uses the normal HST bias
subtraction, flat fielding, and cosmic ray rejection procedures with
the addition of improved pixel interpolation and improved bad/hot
pixel removal.  Information regarding these modifications can be found
online at our web site\footnote{\tt http://etacar.umn.edu} and in a
forthcoming publication (Davidson et al., {\it in preparation\/}).
The spectra are reduced and extracted using approximately the same
parameters used by \citet{heii}.  Each one-dimensional STIS spectrum
discussed here is essentially a 0.1{\arcsec} x 0.25{\arcsec} spatial
sample:  the pixel size is about 0.05\arcsec, the slit width is about
2 CCD columns, each spectral extraction sampled 5 CCD rows, and the
spectral resolution is roughly 52 km s$^{-1}$ at 6307\mbox{\AA}.  We
applied an aperture (extraction height) correction to the absolute
flux, based on an observation of the spectrophotometric standard BD
+75 325 with the same slit and extraction parameters.  Such details
have little effect on the main results of this paper.

In addition to the HST/STIS spectra, we used spectra of the central
star observed with the ESO VLT/UVES.  That observing program is
described in detail by \citet{uveshalpha} and \citet{weis05b}.  Each
one-dimensional UVES spectrum is a 0.30{\arcsec} x 0.91{\arcsec}
spatial sample:  the pixel size is about 0.182{\arcsec}, the slit
width is just under 2 CCD columns, each spectral extraction sampled 5
CCD rows, the seeing ranged from 0.5{\arcsec} to 1.3{\arcsec}, and the
spectral resolution is roughly 3.75 km s$^{-1}$ at 6307\mbox{\AA}.  We
corrected the wavelength scale to the heliocentric reference frame
using the IRAF procedure noao.rvcorrect.  These spectra
are not absolute flux calibrated. 

In ground-based spectra, like those from the VLT/UVES, strong
atmospheric absorption bands and emission lines from the bright ejecta
make it difficult at some wavelengths to detect even dramatic changes
in the spectrum of the central star.  In the VLT/UVES spectra, sharp
lined atmospheric O$_2$ absorption is observed around 6307\mbox{\AA}
and sharp nebular \ion{Fe}{2} $\lambda$6307.04 and \ion{Cr}{2}
$\lambda$6307.39 formed in the nearby ejecta are blended with
the stellar spectra.  The HST/STIS are
unlike ground-based observations in that they are free of atmospheric
absorption and specifically include only the region within $r \approx$
200--300 AU of the central star.  Here ``the star'' or ``$\eta$
Carinae,'' means the central object and its wind, excluding the bright
ejecta and Homunculus nebula (Figure \ref{ejectamap}). If the star is
double then it is unresolved by the HST.  

\subsection{Photometry}
\begin{figure}
\figurenum{5}
\label{nuvevent}
\includegraphics[scale=.75]{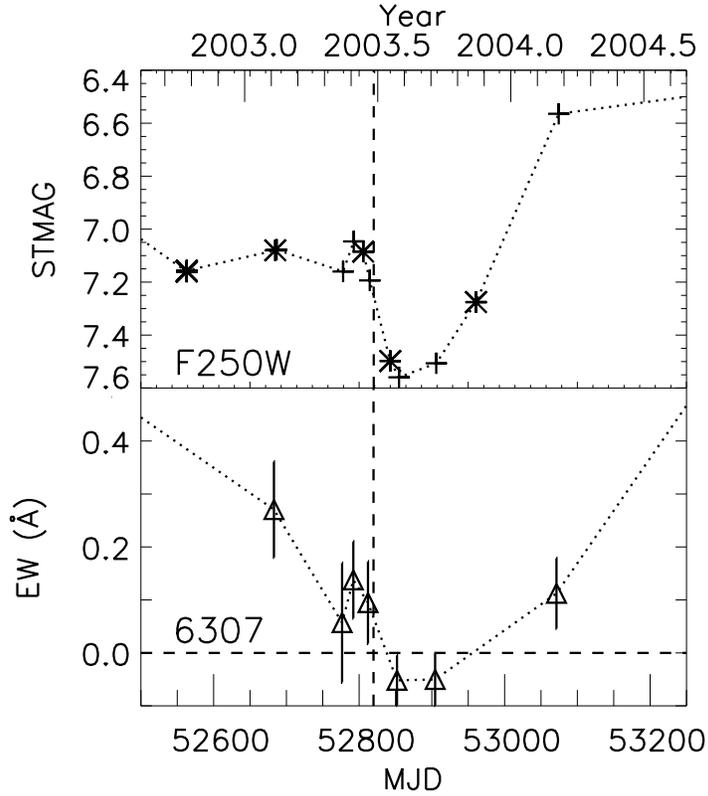}
\caption{Top:  NUV in the F250W filter measured with the ACS/HRC
  (diamonds) and synthesized from flux calibrated STIS/CCD spectra
  (crosses).  The statistical 1$\sigma$ error of each data point is
  0.01--0.02 mag (smaller than the size of the symbols)  Bottom:
  Equivalent width of 6307 {\AA} measured in the STIS/CCD spectra with
  1$\sigma$ error bars estimated from the S/N of the continuum.  The
  vertical dashed line marks the time of the disappearance of the
  \ion{He}{1} line on MJD 52819.8 during the 2003.5 event.}
\end{figure}

Figure \ref{nuvevent} uses HST ACS/HRC data obtained as part of the HST
treasury project.  These data are summarized in Table 3 of
\citet{etaphot}.  The bias-corrected, dark-subtracted, and
flat-fielded images were obtained from the Space Telescope Science
Institute via the Multi-Mission Archive (MAST)
\citep{acscal}\footnote{\tt http://archive.stsci.edu} and measured with
a 0.3$\arcsec$ radius ($\sim$ 10 ACS/HRC pixels) weighted aperture
described by \citet{etaphot}.  The measured flux is corrected to an
infinite aperture using factors we derived following \citet{acscal}
from archived observations of the star GD71.  The corrected fluxes
were converted to the STMAG system \citep{acscal} using the standard
photometry keywords provided by the STScI reduction pipeline in the
FITS headers.  

The ACS/HRC data are supplemented by photometry synthesized from
flux calibrated STIS CCD spectra.  A summary of the spectra 
are given in Table \ref{stisdata}.  They were extracted with a
cross dispersion weighting function which matched the 0.3$\arcsec$
aperture used to measure the ACS/HRC images.  We applied an aperture
(extraction height) correction based on an observation of the
spectrophotometric standard star BD +75 325 with the same slit and
extraction techniques.  The spectra were convolved with the published
ACS/HRC filter and CCD response functions and then integrated to
obtain synthetic fluxes.  Finally, the synthetic fluxes were adjusted
to the STMAG system by comparing the synthetic results to results from
ACS/HRC observations on the same day (MJD 52682). 

\section{Overview of the Feature}
\begin{figure}
\figurenum{2}
\label{6306evol}
\includegraphics[scale=.75]{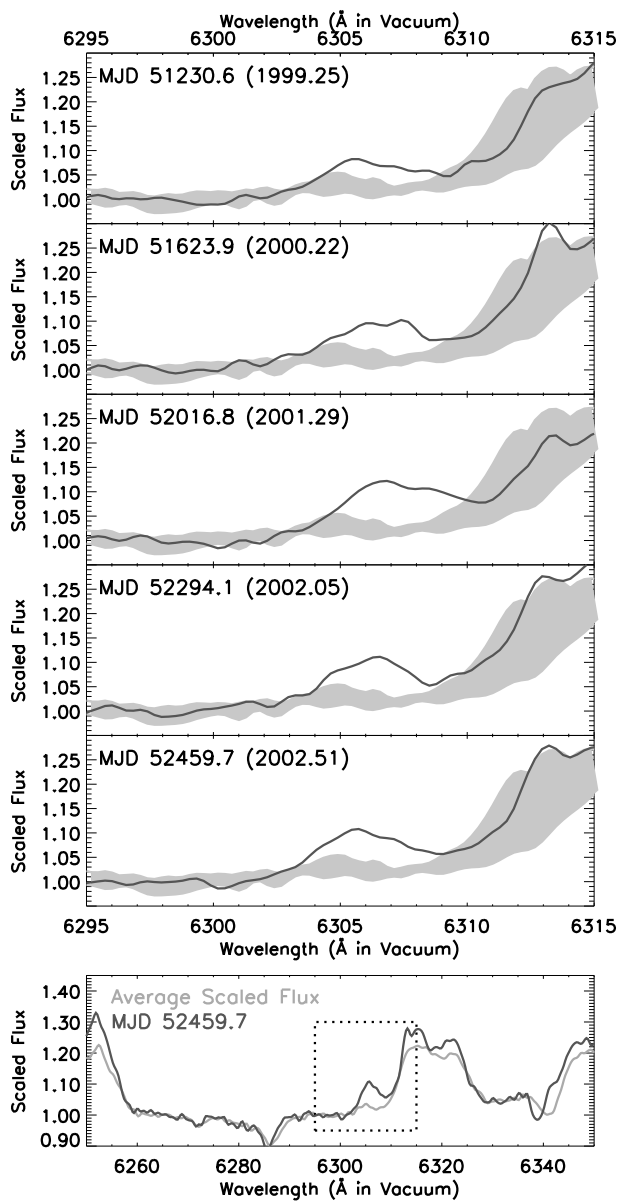}
\caption{Plots of the HST/STIS spectra.  Top five panels:  Plots of
  the 6307\mbox{\AA} emission on the dates that it was unambiguously
  present.  The gray region in each plot gives the range of relative
  flux values when the emission was definitely not present during the
  2003.5 spectroscopic event.  Bottom panel:  A plot of the average
  scaled flux and the flux recorded on MJD 52459.7 over an expanded
  wavelength range.  The dotted box marks the area plotted in the
  previous panels.}   
\end{figure}

\begin{figure}
\figurenum{3}
\label{6306evoluves}
\includegraphics[scale=.75]{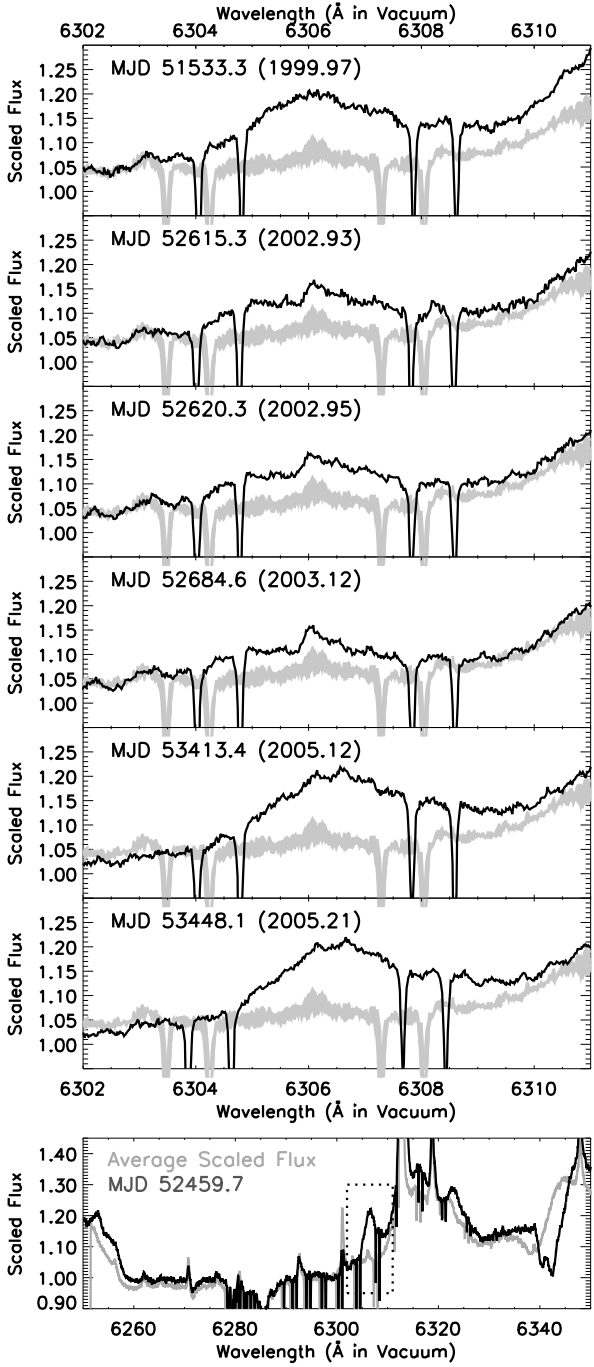}
\caption{Same as for Fig \ref{6306evol} except for the VLT/UVES spectra.}  
\end{figure}

The unidentified emission feature appeared in the wing of the
\ion{Fe}{2} $\lambda$6319 line near 6307\mbox{\AA} (Figure
\ref{6306evol}, Figure \ref{6306evoluves} and Table \ref{tab6307}).
It had a FWHM of about 150 -- 180 km s$^{-1}$ which is narrower than
the stellar wind features (FWHM $\approx$ 300 - 500 km s$^{-1}$) but
significantly broader than the nebular emission from the surrounding
bright ejecta (FWHM $\approx$ 10 km s$^{-1}$
\citep{zethsonphd}\footnote{The spectral resolution of the STIS was
  roughly 52 km s$^{-1}$ at 6307\mbox{\AA}.  Therefore, the
  spectral width of the nebular emission lines is not resolved in the
  STIS/CCD data.}). 

The feature is also present in the spectrum of the star reflected by
the Homunculus lobes (see the VLT/UVES FOS4 slit setting described by
\citet{uveshalpha}).  The equivalent width of the feature is smaller
there.  However there is no variation of its profile along the UVES
slit except for the radial velocity shift introduced by motion of the
reflecting ejecta.  This leads us to conclude that, unlike H$\alpha$
\citep{smith03}, there is no obvious variation of this feature with
stellar latitude. 

\section{Identification}
\begin{figure}
\figurenum{1}
\label{ejectamap}
\includegraphics[scale=2.]{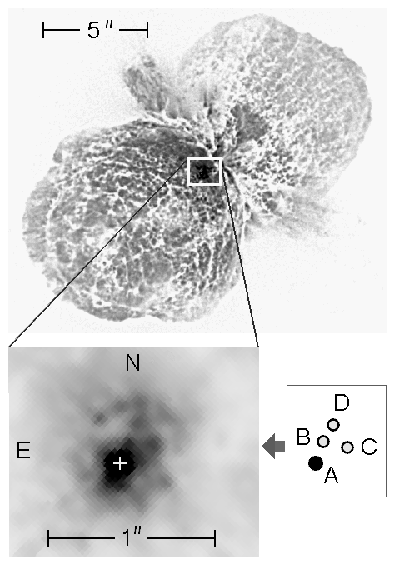}
\caption{Spatial distribution of ejecta mentioned in text.  Top:
   A map of the Homunculus based on HST/WFPC2 images.  The small
   rectangle is the region shown at the bottom.  Lower left:
   HST/ACS image of the inner region, using filter F330W (near UV)
   in 2005.  Lower right:  Relative locations of the slow-moving
   ``Weigelt blobs'' BCD specified by \citet{hw88}, expanded by
   20\% to allow for motions.  The spatial scale throughout this
   figure is adjusted to epoch 2005. }
\end{figure}

We have been unable to find any known transitions consistent with
other lines present in the spectrum that match this feature.  We ruled
out \ion{S}{2} $\lambda$6307 because none of the associated
transitions with similar levels and oscillator strengths at
6314.4\mbox{\AA}, 6288.6\mbox{\AA}, or 6288.0\mbox{\AA} appeared in
the spectrum. Another possible identification may be [\ion{O}{1}]
$\lambda$6302.0 redshifted by 100 -- 200 km s$^{-1}$.  Nearly all the
atomic oxygen in the wind of the central star is probably ionized.
\ion{O}{1} $\lambda$1302, $\lambda$1307, and $\lambda$1306 are present
in the HST/MAMA data but they are {\em blueshifted} by 400 to 500
km/s.  However, the [\ion{O}{1}] $\lambda$6365 is not present and a
100 km s$^{-1}$ redshift would be anomalous.  Therefore, the
6307\mbox{\AA} emission is probably not [\ion{O}{1}] $\lambda$6302.0.
\citet{thackeray}, \citet{damineli1998}, \citet{PASPv113n788}, and
\citet{zethsonphd} reported [\ion{O}{1}] $\lambda$6302.0 in their
spectra, but the line they describe is narrow, redshifted, and
originates in the surrounding bright ejecta, not the central star
(Figure \ref{ejectamap}).  Its wavelength obviously differs from the
feature that we discuss.  

We considered emission pumped by Lyman $\alpha$, Since $\eta$ Car is a
significant source of Lyman $\alpha$ emission, and resonance with that
emission plays a role in the formation of other features
\citep{heii,johansson94,zethsoncrii}.  We found three transitions between 6305
{\AA} -- 6309 {\AA} that are resonant with absorption features within 
3 {\AA} of Lyman $\alpha$:  \ion{Cr}{2} $\lambda$6305.75 (resonant with
$\lambda$1213.499),  \ion{Cr}{3}] $\lambda$6306.8 (resonant with
$\lambda$1215.781), and \ion{Fe}{3}] $\lambda$6306.43 (resonant with
$\lambda$1213.41).  The oscillator strengths have not been measured for
any of these transitions.  While all these species appear in the
nebular spectra of the nearby ejecta, they are not found in the
stellar spectrum.  Furthermore, we do not see these specific
transitions in the nebular spectra along with other previously
identified resonance type features.  Altogether it is difficult to judge
the likelihood of these possible identifications.  However, there is
some appeal to identifying it as an ionized metal line, since aspects
of its variability are similar to those exhibited by the broad
components of the \ion{Fe}{2} lines and the metal absorption
forest around 2500{\AA} (see next section).  

There is an emission line in the spectrum of the surrounding ejecta and
Weigelt Knots at 6306.3 \mbox{\AA} which \citet{zethsonphd} identified
as \ion{Fe}{2} $\lambda$6307.04 and \ion{Cr}{2} $\lambda$6307.39
(Table \ref{zethson}). That line in the spectrum of the ejecta has a
much sharper profile (FWHM $\approx$ 10 km s$^{-1}$) than
6307\mbox{\AA} in the spectrum of the star.  \ion{Fe}{2}
$\lambda$6307.0 and \ion{Cr}{2} $\lambda$6307.4 are also clearly
formed in the area of extended emission within half an arc second of
the star that is resolved by the HST/STIS.  They are not found in the
spectrum of the star itself, whereas the region emitting
6307\mbox{\AA} is spatially unresolved from the central star (Figure
\ref{ejectamap}). 

\section{Variability of the Feature}
\begin{figure}
\figurenum{4}
\label{6306vstime}
\includegraphics[scale=.75]{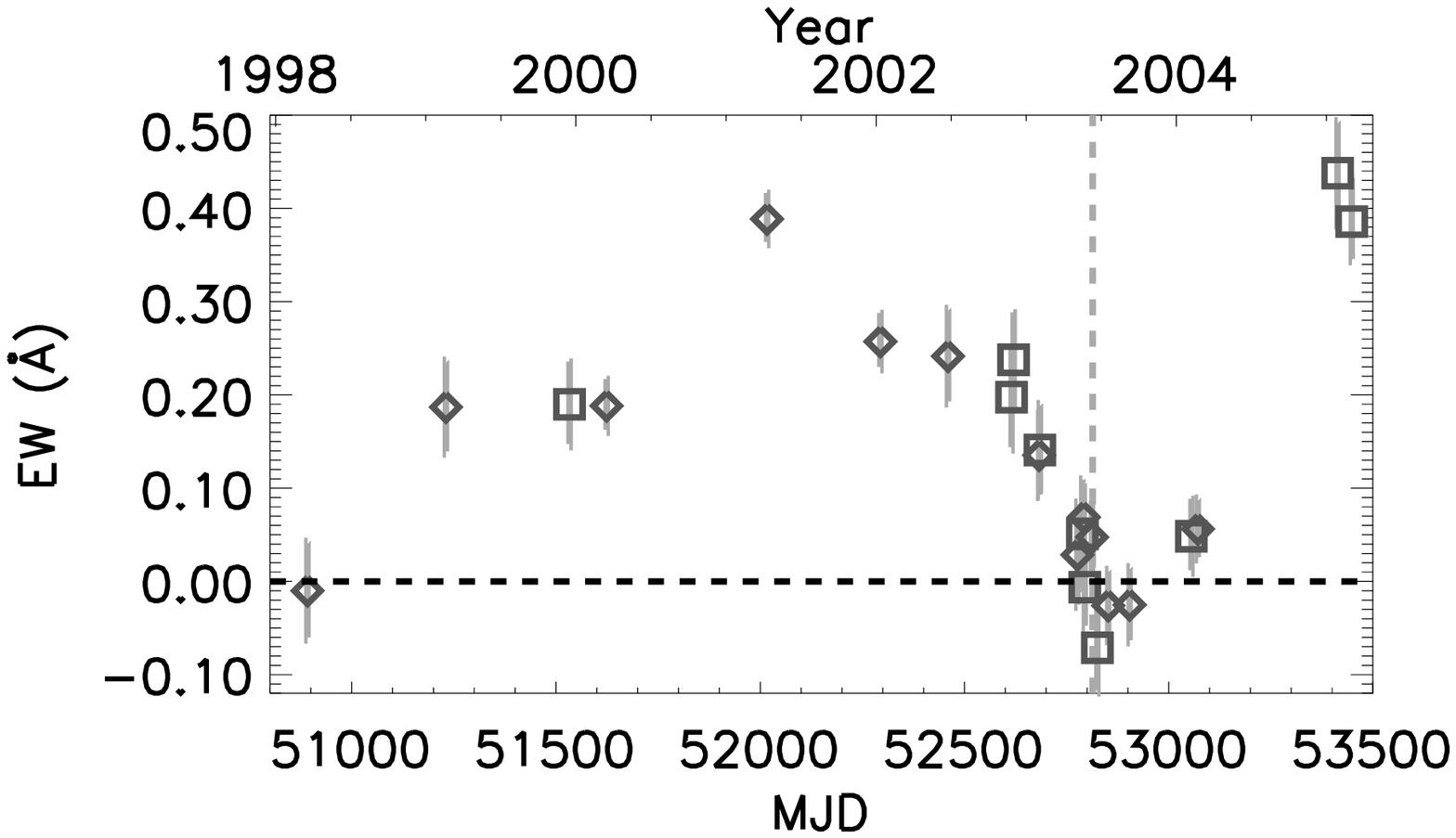}
\caption{The equivalent width of 6307\mbox{\AA} emission in the
  HST/STIS data (diamonds) and VLT/UVES (boxes) in the spectrum of 
  the central star versus time.  The vertical ticks on each point are
  1$\sigma$ error bars given in Table \ref{tab6307}.  The
  dotted vertical line marks the time of the 2003.5 spectroscopic
  event.} 
\end{figure}

The 6307\mbox{\AA} feature {\em disappeared} during the 1998.0 and
2003.5 spectroscopic events (Figure \ref{6306vstime}) at the same time
when the Hydrogen Balmer absorption strengthened and broad high
excitation emission (such as \ion{He}{1}) weakened.  It gradually
declined in flux over the six months prior to the 2003.5 event but
just before disappearing completely its decline was interrupted by a
brief upward tick in brightness lasting a few weeks (Figure
\ref{nuvevent}).  The only other component of the spectrum exhibiting
similar behavior just preceding the event is the ``iron curtain'' of
blanketed near-ultraviolet (NUV) absorption.  One plausible
explanation for this brief ``hiccup'' is a sudden change in ionization
caused by a increase in the ionizing flux from the central star.  In
that case, 6307\mbox{\AA} is a metal line whose population is markedly
increased as the species in the ``iron curtain'' are further ionized
and temporarily depleted.  

\begin{figure}
\figurenum{6}
\label{feiiflux}
\includegraphics[scale=.75]{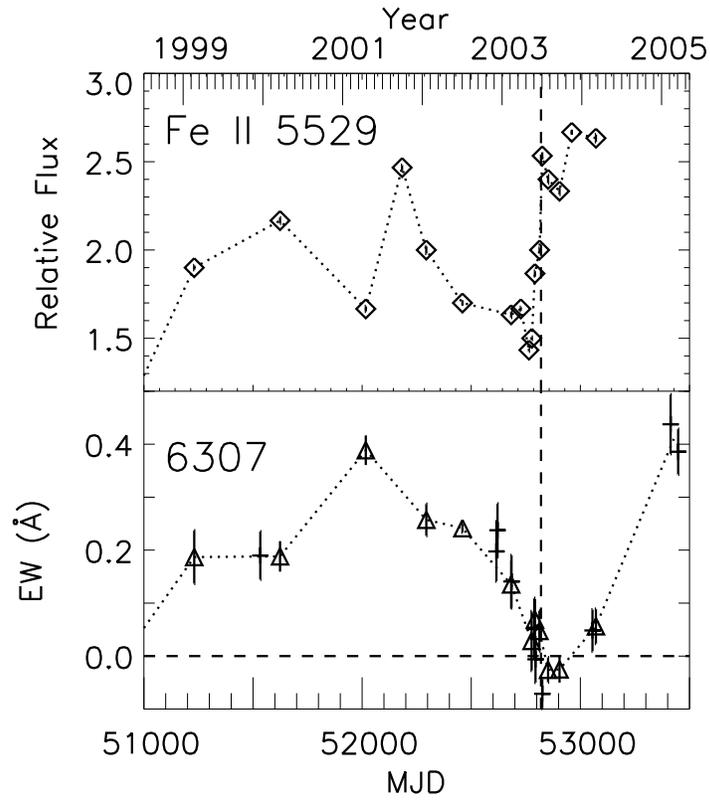}
\caption{Top:  \ion{Fe}{2} 5529 flux (continuum subtracted) measured
  in STIS/CCD spectra.  The 1$\sigma$ error in these data are about
  the size of the plotted symbols.  Bottom:  Equivalent width of 6307 {\AA}
  measured in STIS/CCD spectra (triangles) and VLT/UVES spectra
  (crosses) with 1$\sigma$ error bars.  The vertical dashed line marks
  the time of the 2003.5 event.}
\end{figure}

The 6307 {\AA} feature also evolved much more than most other lines
between the spectroscopic events.  A similar degree of activity is
observed in the broad components of the \ion{Fe}{2} lines formed in
the central star's wind (Figure \ref{feiiflux}).  Note that around
2001 (at a time between spectroscopic events), 6307 {\AA} is anti-correlated
with \ion{Fe}{2} emission.   Like the NUV flux, a rise in 6307 {\AA}
brightness was anti-correlated with \ion{Fe}{2} emission.

The aspects of variability which 6307 {\AA} shares with \ion{Fe}{2}
indicate that it is probably also an ionized metal line formed in the
stellar wind.  The most viable candidates are the transitions we noted
as being in resonance with Lyman $\alpha$ (\ion{Cr}{2}
$\lambda$6305.75, \ion{Cr}{3}] $\lambda$6306.8, and \ion{Fe}{3}]
    $\lambda$6306.43).  For obvious reasons, \ion{Fe}{3}]
      $\lambda$6306.43 is the most enticing option. Unfortunately, the
      atomic data for that transition is lacking so that we are unable
      to confirm our suspicions. 

\section{Mid-Cycle Phenomena:  Evidence for Extra-Cyclical Processes?}

\begin{figure}
\figurenum{7}
\label{orbitfig}
\includegraphics[scale=.40]{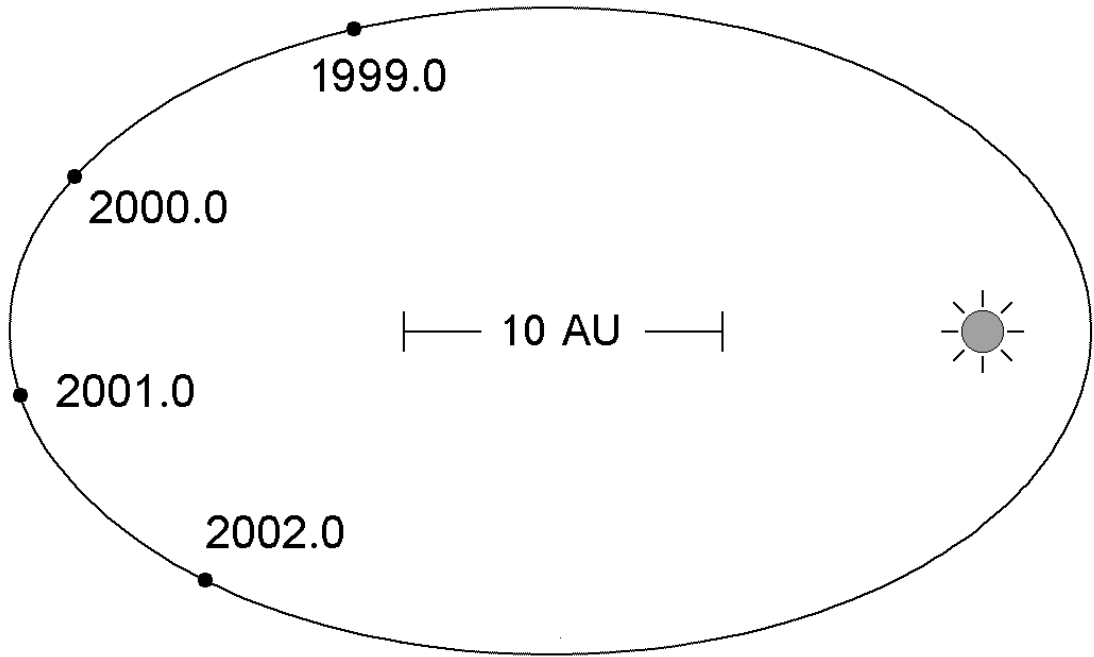}
\caption{Between 1999.0 and 2002.0 the when the 6306 {\AA} feature was
  still varying the secondary was well separated from the primary.
  The time of periasteron is uncertain by several weeks but has little
  effect on this sketch.} 
\end{figure}

In most proposed explanations of the 5.5-year spectroscopic period
{\em there is no obvious reason to expect much variability in
  mid-cycle}, e.g., during 1999--2001 halfway between the 1998.0 and
2003.5 spectroscopic events.  In the most popular scenario describing
these events, the cycle is regulated by a companion star in a highly
eccentric orbit as sketched in Figure \ref{orbitfig}.   At distances
of 20--30 AU from the primary star, where the hypothetical companion
should have been during 1999--2001, relevant wind densities are
factors of 30 to 100  smaller than at periastron.  Column densities
are correspondingly small and the motion is quite slow.  Therefore we
do not expect orbital motion alone to precipitate appreciable
spectroscopic changes during that part of the cycle.  Analogous
comments can be made if the 5.5-year  period is a single-star
thermal/rotational recovery cycle between  outbursts, although such
models are admittedly less definite.  For these reasons, any rapid or
pronounced changes observed in 1999--2001 were most likely {\it not\/}
due to the 5.5-year cycle; instead they probably give us information
about LBV-like random fluctuations in the stellar wind.  If this
statement is wrong, then the mid-cycle variations reveal an aspect of
the cycle that has no explanation in any proposed models.  In either
case it is important to study the features that did vary then.  Among
them the unidentified 6307 {\AA} varied most strongly.  

In our HST/STIS data during the 1998--2003 cycle this line was
brightest in 2001.29 (MJD 52016.8).  Our temporal sampling was too
sparse to indicate the true peak, but on that occasion the line was
more than twice as strong as it had been in 1999-2000 (Figure
\ref{6306vstime}).

Meanwhile, several other changes occurred in 2001, about the same time
as the maximum 6307 {\AA} strength:
\begin{itemize}
\item{A 0.05 magnitude dip in J, H, and K brightness \citep{whitelock},} 
\item{A roughly 20\% increase in near-ultraviolet flux around
  1800\mbox{\AA} over the course of nine months,}
\item{A 15\%--20\% fluctuation in flux emitted in the broad components of
  the \ion{Fe}{2} lines formed in the stellar wind over the course of
  nine to twelve months (see Fig. \ref{feiiflux}),} 
\item{An 20\%--30\% decrease in Hydrogen Balmer P-Cygni absorption
  \citep{halpha},}
\item{A 10\%--20\% increase in the total Hydrogen Balmer emission flux
  \citep{halpha},}  
\item{A sudden and significant increase in the strength of the
  narrow $-140$ km s$^{-1}$ absorption feature in the Hydrogen Balmer lines.} 
\end{itemize}
While these events were simultaneous, we have no proof that they are
physically related.  There is no clear reason or explanation why
any feature should undergo significant change in the {\em middle} of
the spectroscopic cycle.  As noted, in almost any binary star model
of the 5.5-year cycle, the two components were far apart and moving
slowly in 2001.  The moderate changes listed above may perhaps be
ascribed to ordinary LBV-like fluctuations.  Although taken all
together at the same time they are suggestive of some other process at
work in addition to the spectroscopic cycle.  This aspect of the
overall problem of mid-cycle behavior has received little attention to
date.

After the maximum, the emission feature slowly faded over
the next two years until it disappeared completely in the lead up to
the 2003.5 event.  After the event, the feature recovered.  However
on MJD 53413 and MJD 53448 the flux of the feature was more than twice
the flux it had at the same phase a cycle earlier
(approximately MJD 51500).  The fact that the behavior of the feature
appears to be different from cycle to cycle suggests that it is
affected by some additional parameter, i.e. the long-term brightening
trend of the central star.

\section{Summary}
We have discovered a previously unidentified emission feature in
the spectrum of $\eta$ Carinae:  a single emission line at
6307\mbox{\AA}.  This feature is important because: 
\begin{itemize}
\item{The visual light spectrum of $\eta$ Carinae has been extensively
  studied and only about 3\% of the features remain unidentified
  \citep{zethsonphd}.}
\item{This feature's variability appears to be associated with the
  variability of \ion{Fe}{2} in the wind of the central star.} 
\item{It evolves in a unique way with time and thus may
  provide clues to the nature of the spectroscopic cycle and/or long-term
  brightening trend.} 
\item{The peak in the flux of the feature coincided with several
  other mid-spectroscopic cycle changes in the spectrum that we
  cannot explain.} 
\item{The 6307\mbox{\AA} feature is associated with the
  spectroscopic cycle, but its behavior does not reproduce
  exactly from one cycle to the next.}
\end{itemize}

This feature is visible in some spectra of the central
star and its wind, but not others during the last seven years.
We are unable to match it to any published atomic or molecular
transitions of species which we expect to find in the spectrum based
on other lines present.  Its unidentified and transient nature make
it fairly unique in the spectrum of $\eta$ Carinae.  Its disappearance
and reappearance correlated with the spectroscopic events, implying
that it is associated with the spectroscopic cycle.  Its mid-cycle
peak and cycle-to-cycle changes are also sufficiently different from
other identified features so that it deserves special attention.

The 6307\mbox{\AA} feature is visible in ground-based spectra despite
being blended with atmospheric O$_2$ absorption and emission from the
surrounding ejecta.  We encourage our colleagues in the southern
hemisphere to look for it since it varies between spectroscopic events
and tracking these variations with better temporal sampling may help
provide further insight into the 5.5-year cycle or the recent dramatic
brightening of the central star. 

\section{Acknowledgments}
This work made use of the NIST Atomic Spectra
Database\footnote{http://physics.nist.gov/PhysRefData/ASD/index.html}
and the Kentucky Atomic Line List
v2.04\footnote{http://www.pa.uky.edu/$\sim$peter/atomic/index.html}. 
We also wish to thank M. Salvo (ANU) for generously using some of her
time on the MSO 2.3m to obtain current ground-based spectra for us.
T.R. Gull (NASA/GSFC) prepared most of the detailed STIS observing
plans, gave other valuable help in the Treasury Program.  Meanwhile,
K. Ishibashi (MIT) produced the improved reduction software and
contributed to the observing plans.  We also thank S Johansson, H
Hartman (University of Lund), and M. Bautista (Inst. Venezolano
Invest. Cientifica) for comments on an early version of this paper,
and Beth Periello (STScI) for assistance with the HST observing plan.

\clearpage

\begin{deluxetable}{lllrrr}
\tablewidth{0pt}
\tabletypesize{\small}
\tablecolumns{5}
\tablecaption{Narrow Nebular Emission Lines Identified By Zethson Near 6307{\AA}\tablenotemark{a}\label{zethson}}
\tablehead{\colhead{$\lambda_{obs}$ {\AA}}&\colhead{Species}&
  \colhead{Transition}&\colhead{$\lambda_{lab}$}&\colhead{Notes\tablenotemark{b}}} 
\startdata
6256.56&\ion{Fe}{2}(34)&$b^4F_{3/2}-z^6D_{3/2}$&6257.08&\nodata\\
6261.88&[\ion{Fe}{2}](44F)&$a^2G_{9/2}-b^2F_{5/2}$&6265.93&\nodata\\
6265.17&[\ion{V}{2}]&$a^3D_1-b^1F_3$&6265.93&id?\\
\nodata&[\ion{V}{2}]&$b^3G_3-d^3P_1$&6266.33&id?\\
\nodata&[\ion{V}{2}]&$a^3G_5-d^3F_4$&6266.34&id?\\
6270.62&\ion{Fe}{2}&$b^2H_{11/2}-z^6P_{3/2}$&6271.70&\nodata\\
6276.44&[\ion{Fe}{2}](15F)&$a^4F_{7/2}-a^2P_{3/2}$&6277.2&\nodata\\
6280.81&[\ion{Fe}{2}]&$a^2D2_{3/2}-a^2S_{1/2}$&6281.69&\nodata\\
\nodata&\ion{Fe}{2}]&$b^4F_{5/2}-z^6D_{5/2}$&6281.56&\nodata\\
6287.00&[\ion{Mn}{2}]&$a^5D_2-a^3P_1$&6287.77&\nodata\\
6292.65&\ion{Fe}{2}&$(^5D)4d^4P_{3/2}-(^5D_0)4f^2[3]_{5/2}$&6293.57&?E$_u$\\
6297.32&Unidentified&\nodata&\nodata&\nodata\\
6301.14&[\ion{O}{1}](1F)&$2p^4~^3P_2-2p^4~^1D_2$&6302.05&\nodata\\
6306.3&\ion{Fe}{2}(200)&$c^4F_{9/2}-x^4F_{9/2}$&6307.47&\nodata\\
\nodata&\ion{Cr}{2}&$b^2P_{3/2}-y^2P_{3/2}$&6307.39&\nodata\\
6308.51&\ion{Fe}{2}(34)&$b^4F_{7/2}-z^6D_{7/2}$&6309.27&\nodata\\
6312.82&[\ion{S}{3}](3F)&$3p^2~^1D_2-3p^2~^1S_0$&6313.81&Not in 98\\
6318.77&\ion{Fe}{2}&$z^4D_{7/2}-c^4D_{7/2}$&6319.73&4p-4s\\
6347.95&\ion{Si}{2}(2)&$4s~^2S_{1/2}-4p~^2P_{3/2}$&6348.84&\nodata\\
6357.83&\ion{Fe}{2}&$(^5D)4d~^4P_{5/2}-(^5D_2)4f^2[4]_{7/2}$&6358.92&?E$_u$\\
\nodata&[\ion{Mn}{2}]&$a^5D_0-a^3P_1$&6359.21&\nodata\\
\enddata
\tablenotetext{a}{From \citet{zethsonphd}.}
\tablenotetext{b}{Taken from the ``Notes'' column of Zethson's tables:
  id? = uncertain identification; ?E$_u$ = \ion{Fe}{2} and \ion{Cr}{2}
  transitions with an upper level $\ge$ 10 eV and the excitation
  mechanism is ``questionable'';  Not in 98 = Not present in the spectrum
  observed just after the 1998.0 spectroscopic event; 4p-4s = a 4p-4s
  transition.} 
\end{deluxetable}

\begin{deluxetable}{lrrrrrr}
\tablewidth{0pt}
\tabletypesize{\tiny}
\tablecolumns{7}
\tablecaption{HST STIS Data\label{stisdata}}
\tablehead{
\colhead{Root}&&&\colhead{Slit Angle}&&\colhead{Central
  $\lambda$}&\colhead{Exp Length}\\
\colhead{Name}&\colhead{MJD}&\colhead{Slit}&\colhead{(deg)\tablenotemark{a}}
&\colhead{Grating}&\colhead{({\AA})}&\colhead{(sec)}}
\startdata
\multicolumn{7}{c}{6307{\AA} Observations}\\
\hline
o4j801120&50891.7&52x0.1&-28.&G750M&6252&9.4\\
o556020p0&51230.6&52x0.1&-28.&G750M&6252&8.0\\
o5kz010m0&51623.9&52x0.1&-28.&G750M&6252&10.0\\
o62r010l0&52016.8&52x0.1&+22.&G750M&6252&10.0\\
o6ex020l0&52294.1&52x0.1&-82.&G750M&6252&8.0\\
o6mo02150&52459.7&52x0.1&+69.&G750M&6252&8.0\\
o8gm12050&52682.9&52x0.1&-57.&G750M&6252&8.0\\
o8gm330r0&52776.6&52x0.1&+38.&G750M&6252&8.0\\
o8gm521s0&52792.0&52x0.1&+62.&G750M&6252&8.0\\
o8gm520h0&52812.2&52x0.1&+70.&G750M&6252&15.0\\
o8ma820a0&52852.0&52x0.1&+105.&G750M&6252&8.0\\
o8ma920z0&52904.5&52x0.1&+153.&G750M&6252&8.0\\
o8ma940r0&53071.3&52x0.1&-28.&G750M&6252&10.0\\
\hline
\multicolumn{7}{c}{\ion{Fe}{2} $\lambda$5529 Observations}\\
\hline
o4j8010d0&50891.5&52x0.1&-28&G750M&5734&15.0\\
o55602090&51230.5&52x0.1&-28&G750M&5734&15.0\\
o62r01090&52016.8&52x0.1&+21&G750M&5734&8.0\\
o6ex030b0&52183.1&52x0.1&+165&G750M&5734&15.0\\
o6ex02090&52294.0&52x0.1&-82&G750M&5734&6.0\\
o6mo020i0&52459.6&52x0.1&+69&G750M&5734&9.0\\
o8gm120a0&52682.9&52x0.1&-57&G750M&5734&6.0\\
o8gm210d0&52727.3&52x0.1&-28&G750M&5734&9.0\\
o8gm410d0&52764.4&52x0.1&+27&G750M&5734&9.0\\
o8gm320i0&52778.6&52x0.1&+38&G750M&5734&9.0\\
o8gm520i0&52791.8&52x0.1&+62&G750M&5734&9.0\\
o8gm620i0&52814.0&52x0.1&+70&G750M&5734&9.0\\
o8ma720h0&52825.4&52x0.1&+69&G750M&5734&9.0\\
o8ma820j0&52852.0&52x0.1&+105&G750M&5734&6.0\\
o8ma920b0&52904.4&52x0.1&+153&G750M&5734&6.0\\
o8ma830d0&52960.7&52x0.1&-142&G750M&5734&8.0\\
o8ma94080&53071.3&52x0.1&-28&G750M&5734&6.0\\
\hline
\multicolumn{7}{c}{Spectra Used to Synthesize ACS/HRC F250W Fluxes}\\
\hline
o8gm12030&52682.9&52x0.1&-57.&G230MB&2836&300.0\\
o8gm12090&52682.9&52x0.1&-57.&G230MB&2557&800.0\\
o8gm120b0&52682.9&52x0.1&-57.&G430M&3680&52.0\\
o8gm120c0&52682.9&52x0.1&-57.&G230MB&2697&340.0\\
o8gm120d0&52682.9&52x0.1&-57.&G430M&3423&90.0\\
o8gm120h0&52683.0&52x0.1&-57.&G230MB&1995&600.0\\
o8gm120l0&52683.0&52x0.1&-57.&G230MB&2135&600.0\\
o8gm120n0&52683.0&52x0.1&-57.&G430M&3165&90.0\\
o8gm120r0&52683.0&52x0.1&-57.&G230MB&2416&320.0\\
o8gm120t0&52683.0&52x0.1&-57.&G230MB&2976&340.0\\
o8gm120w0&52683.0&52x0.1&-57.&G230MB&2276&600.0\\
o8gm33020&52776.4&52x0.1&+38.&G230MB&2135&300.0\\
o8gm33060&52776.4&52x0.1&+38.&G430M&3165&90.0\\
o8gm33090&52776.5&52x0.1&+38.&G230MB&3115&300.0\\
o8gm330e0&52776.5&52x0.1&+38.&G230MB&2416&320.0\\
o8gm330i0&52776.5&52x0.1&+38.&G230MB&2976&320.0\\
o8gm330n0&52776.5&52x0.1&+38.&G230MB&2276&300.0\\
o8gm32050&52776.6&52x0.1&+38.&G230MB&2836&300.0\\
o8gm320h0&52778.6&52x0.1&+38.&G230MB&2557&400.0\\
o8gm320l0&52778.7&52x0.1&+38.&G430M&3680&52.0\\
o8gm320m0&52778.7&52x0.1&+38.&G230MB&2697&340.0\\
o8gm320p0&52778.7&52x0.1&+38.&G430M&3423&90.0\\
o8gm320x0&52778.7&52x0.1&+38.&G230MB&1995&300.0\\
o8gm52050&52791.7&52x0.1&+62.&G230MB&2836&300.0\\
o8gm520h0&52791.8&52x0.1&+62.&G230MB&2557&400.0\\
o8gm520l0&52791.8&52x0.1&+62.&G430M&3680&52.0\\
o8gm520m0&52791.8&52x0.1&+62.&G230MB&2697&340.0\\
o8gm520p0&52791.8&52x0.1&+62.&G430M&3680&90.0\\
o8gm520x0&52791.8&52x0.1&+62.&G230MB&1995&300.0\\
o8gm52100&52791.9&52x0.1&+62.&G230MB&2135&400.0\\
o8gm52170&52791.9&52x0.1&+62.&G430M&3165&90.0\\
o8gm52180&52791.9&52x0.1&+62.&G230MB&3115&300.0\\
o8gm521f0&52791.9&52x0.1&+62.&G230MB&2416&320.0\\
o8gm521j0&52791.9&52x0.1&+62.&G230MB&2976&340.0\\
o8gm521o0&52792.0&52x0.1&+62.&G230MB&2276&300.0\\
o8gm63040&52812.1&52x0.1&+70.&G230MB&2416&350.0\\
o8gm63080&52812.2&52x0.1&+70.&G230MB&2976&320.0\\
o8gm630d0&52812.2&52x0.1&+70.&G230MB&2836&300.0\\
o8gm62050&52813.7&52x0.1&+70.&G230MB&2836&300.0\\
o8gm620h0&52814.0&52x0.1&+70.&G230MB&2557&400.0\\
o8gm620l0&52814.1&52x0.1&+70.&G430M&3680&52.0\\
o8gm620m0&52814.1&52x0.1&+70.&G230MB&2697&340.0\\
o8gm620p0&52814.1&52x0.1&+70.&G430M&3423&90.0\\
o8gm620x0&52814.2&52x0.1&+70.&G230MB&1995&300.0\\
o8gm62100&52814.2&52x0.1&+70.&G230MB&2135&300.0\\
o8gm62140&52814.2&52x0.1&+70.&G430M&3165&90.0\\
o8ma82060&52851.9&52x0.1&+105.&G230MB&2836&300.0\\
o8ma820i0&52852.0&52x0.1&+105.&G230MB&2557&400.0\\
o8ma820m0&52852.1&52x0.1&+105.&G430M&3680&52.0\\
o8ma820n0&52852.1&52x0.1&+105.&G230MB&2697&340.0\\
o8ma820q0&52852.1&52x0.1&+105.&G430M&2697&90.0\\
o8ma820y0&52852.2&52x0.1&+105.&G230MB&1995&300.0\\
o8ma82110&52852.2&52x0.1&+105.&G230MB&2135&300.0\\
o8ma821a0&52852.3&52x0.1&+105.&G430M&3165&90.0\\
o8ma821b0&52852.3&52x0.1&+105.&G230MB&2416&320.0\\
o8ma821i0&52852.4&52x0.1&+105.&G230MB&2976&300.0\\
o8ma821m0&52852.4&52x0.1&+105.&G230MB&2276&300.0\\
o8ma92040&52940.3&52x0.1&+153.&G230MB&2836&300.0\\
o8ma920a0&52940.3&52x0.1&+153.&G230MB&2557&800.0\\
o8ma920c0&52940.4&52x0.1&+153.&G430M&3680&52.0\\
o8ma920d0&52940.4&52x0.1&+153.&G230MB&2697&340.0\\
o8ma920e0&52940.4&52x0.1&+153.&G430M&3423&90.0\\
o8ma920i0&52940.4&52x0.1&+153.&G230MB&1995&600.0\\
o8ma920m0&52940.4&52x0.1&+153.&G230MB&2135&600.0\\
o8ma920o0&52940.4&52x0.1&+153.&G430M&3350&90.0\\
o8ma920p0&52940.4&52x0.1&+153.&G230MB&3115&300.0\\
o8ma920s0&52940.5&52x0.1&+153.&G230MB&2416&600.0\\
o8ma920u0&52940.5&52x0.1&+153.&G230MB&2976&340.0\\
o8ma920x0&52940.5&52x0.1&+153.&G230MB&2276&300.0\\
o8ma94020&53071.3&52x0.1&-28.&G230MB&2836&320.0\\
o8ma94070&53071.3&52x0.1&-28.&G230MB&2557&410.0\\
o8ma94090&53071.3&52x0.1&-28.&G430M&2557&52.0\\
o8ma940a0&53071.3&52x0.1&-28.&G430M&3423&90.0\\
o8ma940e0&53071.3&52x0.1&-28.&G230MB&2697&323.0\\
o8ma940h0&53071.3&52x0.1&-28.&G430M&3165&90.0\\
o8ma940i0&53071.3&52x0.1&-28.&G230MB&2135&320.0\\
o8ma940m0&53071.3&52x0.1&-28.&G230MB&2416&450.0\\
\enddata
\tablenotetext{a}{The slit angel is measured from north through east.
  All slits are peaked up on the central star.} 
\end{deluxetable}

\begin{deluxetable}{lllrrr}
\tablewidth{0pt}
\tabletypesize{\small}
\tablecolumns{6}
\tablecaption{Measured Properties of 6307\mbox{\AA}\label{tab6307}}
\tablehead{&&&&\colhead{Line
    Flux\tablenotemark{a}}&\colhead{Line EW\tablenotemark{a}}\\
\colhead{MJD}&\colhead{Year}&\colhead{Telescope}&\colhead{Centroid (\mbox{\AA})}&\colhead{(erg cm$^{-2}$ s$^{-1}$)$\times 10^{-13}$}&\colhead{(\mbox{\AA})}}
\startdata
50891.7&1998.21&HST/STIS&\nodata&2.98$\pm$1.52&-0.01$\pm$0.05\tablenotemark{b}\\
51230.6&1999.14&HST/STIS&6306.88$\pm$0.19&9.77$\pm$2.66&0.19$\pm$0.05\\
51533.3&1999.97&VLT/UVES&6306.80$\pm$0.08&\nodata&0.19$\pm$0.05\tablenotemark{b}\\
51623.9&2000.22&HST/STIS&6306.95$\pm$0.19&8.89$\pm$1.36&0.19$\pm$0.03\\
52016.8&2001.29&HST/STIS&6307.32$\pm$0.19&22.62$\pm$1.63&0.39$\pm$0.03\\
52294.1&2002.05&HST/STIS&6306.90$\pm$0.19&14.91$\pm$1.78&0.26$\pm$0.03\\
52459.7&2002.51&HST/STIS&6306.75$\pm$0.19&14.92$\pm$3.20&0.24$\pm$0.05\\
52615.3&2002.93&VLT/UVES&6306.50$\pm$0.08&\nodata&0.20$\pm$0.06\\
52620.3&2002.95&VLT/UVES&6306.69$\pm$0.08&\nodata&0.24$\pm$0.05\\
52682.9&2003.12&HST/STIS&6306.75$\pm$0.19&7.62$\pm$2.56&0.14$\pm$0.05\\
52684.6&2003.12&VLT/UVES&6306.41$\pm$0.08&\nodata&0.14$\pm$0.05\\
52776.6&2003.24&HST/STIS&\nodata&1.79$\pm$3.56&0.03$\pm$0.06\\
52788.6&2003.40&VLT/UVES&\nodata&\nodata&0.05$\pm$0.06\\
52792.0&2003.42&HST/STIS&\nodata&4.65$\pm$2.46&0.07$\pm$0.04\\
52794.5&2003.42&VLT/UVES&\nodata&\nodata&-0.01$\pm$0.04\\
52812.2&2003.47&HST/STIS&\nodata&3.61$\pm$2.98&0.05$\pm$0.04\\
52825.5&2003.51&VLT/UVES&\nodata&\nodata&-0.07$\pm$0.05\\
52852.0&2003.58&HST/STIS&\nodata&-1.90$\pm$2.87&-0.03$\pm$0.04\\
52904.5&2003.72&HST/STIS&\nodata&-2.19$\pm$3.60&-0.03$\pm$0.04\\
53055.6&2004.14&VLT/UVES&\nodata&\nodata&0.05$\pm$0.04\\
53071.3&2004.18&HST/STIS&\nodata&5.41$\pm$3.22&0.06$\pm$0.03\\
53413.4&2005.12&VLT/UVES&6306.99$\pm$0.08&\nodata&0.44$\pm$0.06\\
53448.1&2005.21&VLT/UVES&6306.93$\pm$0.08&\nodata&0.39$\pm$0.04\\
\enddata
\tablenotetext{a}{Fluxes and equivalent widths are expressed relative
  to the continuum with values greater than zero denoting excess flux
  above the continuum.  Only the STIS/CCD data is absolute flux calibrated.}
\tablenotetext{b}{The wing of \ion{Fe}{2} $\lambda$ 6319 was brighter than 
  normal at this time so that it interfered with the measurement of
  the equivalent width of the feature.}
\end{deluxetable}

\end{document}